# Atomic process of oxidative etching in monolayer molybdenum disulfide

Danhui Lv,[a] Hulian Wang,[a] Dancheng Zhu,[a] Jie Lin,[a] Guoli Yin,[a, #] Fang Lin[b], Ze Zhang,[a] Chuanhong Jin[a,*]

[a] State Key Laboratory of Silicon Materials, School of Materials Science and Engineering, Zhejiang University, Hangzhou, Zhejiang 310027, P.R. China

[b] College of Electronic Engineering, South China Agricultural University, Guangzhou, Guangdong 510642, P.R. China

[#] Present address: Department of Materials Science and Engineering, Stanford University, Stanford, CA 94305, USA

[*] Corresponding author: Chuanhong Jin, E-mail: chhjin@zju.edu.cn, Tel & Fax: +86-571-87953700.

**Abstract**

The microscopic process of oxidative etching of two-dimensional molybdenum disulfide (2D $MoS_2$) at an atomic scale is investigated using a correlative TEM-etching study. $MoS_2$ flakes on graphene TEM grids are precisely tracked and characterized by TEM before and after the oxidative etching. This allows us to determine the structural change with an atomic resolution on the edges of the domains, of well-oriented triangular pits and along the grain boundaries. We observe that the etching mostly starts from the open edges, grain boundaries and pre-existing atomic defects. A zigzag Mo edge is assigned as the dominant termination of the triangular pits, and profound terraces and grooves are observed on the etched edges. Based on the statistical TEM analysis, we reveal possible routes for the kinetics of the oxidative etching in 2D $MoS_2$, which should also be applicable for other 2D transition metal dichalcogenide materials like $MoSe_2$ and $WS_2$.

## 1. Introduction

Atomically thin transition metal dichalcogenide (TMD) materials, as an emerging family of 2D materials, have been extensively studied in recent years. Their unique polymorph phase structures exhibit excellent properties with promising applications in electronics, optoelectronics and catalysis.[1-10] To further tune the structures and associated properties of these 2D TMD materials, a number of methods have been proposed, such as laser illumination, lithography patterning and so forth.[11-14] Of these methods, oxidative etching, by simply heating the sample in air, is efficient to create well-oriented triangular pits and further decrease the thickness of 2D TMD materials. The extra edge states introduced by these pits (sometimes also called antidots) will tailor the electronic structures, and enhance the catalytic properties of 2D $MoS_2$ for hydrodesulphurization (HDS) reactions and hydrogen evolution reduction (HER) reactions[15-16], as the etching is initiated preferably from defective sites[17-21]. Nonetheless, oxidative etching can also be used for visualizing grain boundaries and atomic vacancies in 2D $MoS_2$ films upon heating in air, and UV illumination in the presence of $O_3$ and/or silver nanoparticles[22-24].

At the relatively lower temperature regime, oxidation and the subsequent aging effect have become critical issues for 2D TMD materials towards its practical applications, and they exhibit poor air stability with a degrading performance upon long-term air exposure. Microscopically, chemically adsorbed oxygen onto monolayer $MoS_2$ causes its carrier type to change from n-type to p-type in a $MoS_2$-based field effect transistor, owing to charge doping[25]. Furthermore, the atomically absorbed oxygen is also regarded as the major contributor to the substantially enhanced photoluminescence in monolayer $MoS_2$[20]. Recently, Yamamoto et al.[26] demonstrated that the oxidization of the top-few layers in tungsten diselenide flakes will provide good dielectric layers for device fabrication and passivation.

Currently, there have been a number of microscopic mechanisms proposed to clarify the roles played by a few important parameters and the associated kinetics for the oxidative etching in 2D $MoS_2$. For instance, the anisotropic etching mechanism can explain the formation of well-oriented triangle pits with zigzag edges [19-20], although the assignment of edge structures is still controversial (either ZZ-Mo or ZZ-S). Moreover, the role played by oxygen and its interaction with exiting atomic detects has been addressed theoretically [27-28], while the atomic process for the reaction pathway remains unclear. Hence, whether any atomic or cluster forms of oxygen resides in the oxides is inconsistent among different reports. All these issues are essential to reveal the microscopic mechanisms and kinetics of oxidative etching and other oxidation behavior in 2D TMD materials, and the subsequent tunability of structures and properties. Herein, we carry out a correlative oxidative etching-STEM study to reveal the kinetics for oxidative etching in monolayer $MoS_2$. By directly comparing the structural evolution of the edges, triangle pits and grain boundaries of

MoS$_2$ flakes, we propose a possible route for the microscopic reaction pathway during oxidative etching in monolayer MoS$_2$.

## 2. Material and Methods

MoS$_2$ samples were grown on graphene-supported TEM grids, according to the reported method[29]. We conducted STEM characterization on the same MoS$_2$ flake before and after oxidative etching, without any chemistry-based transferring process. Oxidative etching was performed by heating the samples inside a mini-CVD furnace (Lindberg) in air. The furnace was heated up to 300 °C in 10 minutes, and held at this temperature for 5 minutes for the etching. Subsequently, the TEM grids were immediately removed from the furnace for further characterization. ADF-STEM was performed in an aberration-corrected FEI Titan STEM, with ChemiSTEM capability, operated at 200 kV. The convergence semi-angle for the incident electron probe was set to ~22 mrad, and the half-angle for image collection was set to ~43 to ~200 mrad. The electron probe current was ~70 pA, and the dwell time per pixel was ~10 μs during the imaging, corresponding to an electron dose level of ~30 e/nm$^2$. The STEM images, without noticeable structural damage on the oxidized MoS$_2$, were used for further analysis, which were mainly recorded during the first few rounds of beam scanning.

## 3. Results and Discussion

Fig. 1(a) and 1(b) show the ADF-STEM images of the morphology and atomic structure of the as-grown MoS$_2$ flakes on graphene TEM grids, respectively, in which monolayer MoS$_2$ were mostly shaped as truncated triangles with an average edge length of ~80 nm. There were three typical edge terminations observed: the dominant ZZ-Mo edge (marked by the blue line), ZZ-S edge (marked by the red line) and ZZ-S-Mo edge (marked by the green line), as shown in Fig. 1(b) and Fig. S1. Note that the ZZ-S-Mo edge and ZZ-S edge belong to the same crystalline orientation (-1010), and the only difference was that there was a bare Mo atom attached to the outmost sulfur, similar to the Klein edge in graphene[30-31].

Fig. 1(c) and 1(d) show the structure evolution of a MoS$_2$ monolayer before and after oxidative etching. As is clearly seen, the monolayer undergoes a dramatic change in the morphology, from a triangle to an irregular polygon, with considerable mass loss during the etching. Here, the well-oriented triangular pits are rarely observed, which is in contrast to previous reports. This will be discussed in detail below. Local edge terminations of different segments along the edge of the etched MoS$_2$ flake are marked as shown in Fig. 1(e). Profound terraces with atomic steps and local grooves were formed on the edges as a result of oxidative etching.

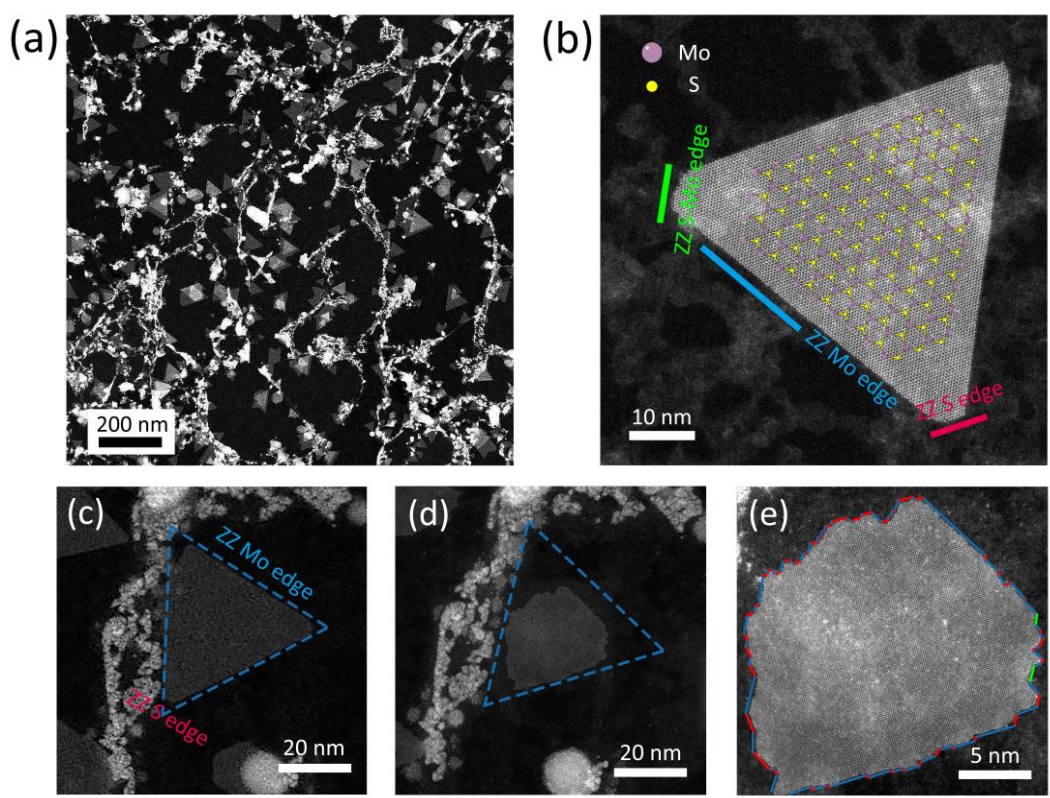

**Fig. 1** ADF-STEM images of MoS$_2$ grown on a graphene substrate. (a) A typical low-magnification ADF-STEM image of MoS$_2$ grown on a graphene substrate. The white region beside MoS$_2$ are MoO$_x$ residues from the CVD growth and PMMA residues from the graphene transfer. (b) Atomic resolution STEM image of single domain MoS$_2$ grown on a graphene substrate. The hexagonal atomic model indicates the edge structures (purple: Mo, yellow: S) which are also marked on the real structure by colored lines (blue line: zigzag Mo edge, red line: zigzag S edge, green line: zigzag S-Mo edge). ADF-STEM images of a MoS$_2$ single domain before (c) and after (d) 10 minutes annealing in air at 300 ℃. The dotted triangles in both images mark the MoS$_2$ domain before etching, and (e) different edges labelled by different colors showing the MoS$_2$ domain after etching (blue: zigzag Mo edge, red: zigzag S edge, green: zigzag S-Mo edge).

Fig. 2 shows the atomic structures along a triangular pit formed on monolayer $MoS_2$ after etching, which has rarely been observed under the present experimental conditions. A nearly equilateral triangular shape with an edge length of 10 nm is obtained. By assigning the edge terminations and marking them with different colored lines, we observed that the whole edge line was comprised primarily of ZZ-Mo edges (marked by blue lines), rather than the ZZ-S and ZZ-S-Mo edges (marked by red and green lines, respectively). Similar results were observed on all the triangle pits formed on the $MoS_2$ flakes grown on either graphene substrates or the conventional $SiO_2$/Si substrates (please refer to Fig. S2–S5 for more details). Under the same conditions for oxidative etching, ZZ-terminated Mo atoms seemed to be thermodynamically more stable than ZZ-terminated S atoms, which resulted in the formation of ZZ-Mo dominant terminations on the pits. This explanation is also consistent with previous studies that reported that oxidation was preferably initiated from the pre-existing atomic S vacancies inside a domain or the under-saturated S atoms on the edge, as they could facilitate the adsorption and dissociation of oxygen molecules, and thus promoted the oxidation reaction[32]. The less-common triangular pits on oxidative etched $MoS_2$ flakes on graphene, which we observed, further supported this explanation. During our ADF-STEM imaging, very few sulfur vacancies were observed on the as-grown $MoS_2$, except for those caused by beam damage during the STEM characterizations. As such, it became difficult for oxidation to occur at the interior single-crystalline domain of the high-quality $MoS_2$ monolayer. This was also consistent with previous studies that reported $MoS_2$ grown on graphene substrate is usually of high quality.

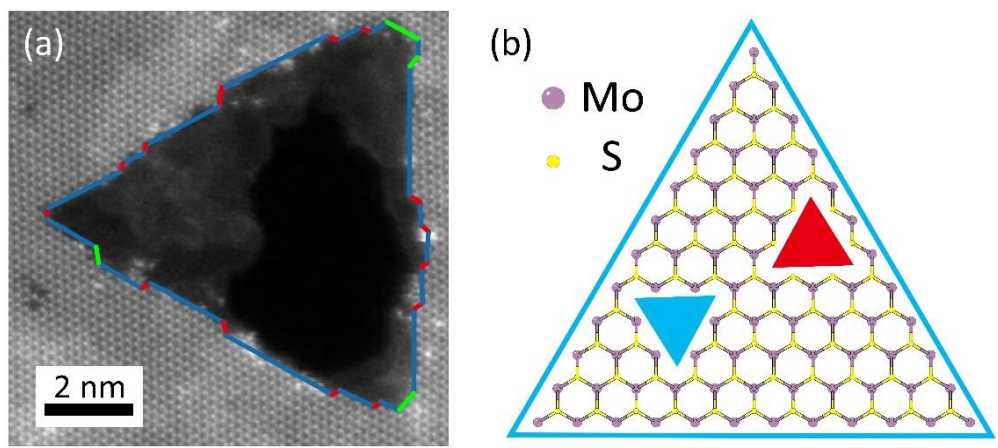

**Fig. 2** ADF-STEM images of etched inner domains in as-grown MoS$_2$ monolayers. (a) ADF-STEM image of etched triangular pits in an as-grown MoS$_2$ monolayer. (b) Schematic atomic model showing the two kinds of etching edges (red for zigzag Mo edge, blue for zigzag S edge) in a zigzag Mo edge MoS$_2$ domain. Only zigzag Mo edge predominant etching pits were observed in our experiment.

To determine the possible atomic pathway for the oxidative etching, we performed a detailed analysis on the evolution of the edge structures at an atomic scale over a number of oxidative etched MoS$_2$ flakes. Two representative examples are given in Fig. 3(a) and 3(d) for the ZZ-S/ZZ-S-Mo edge and ZZ-Mo edge, respectively, where the local edge structures were assigned and then marked with different colored lines. A general view shows that terraces with atomic steps were frequently formed on the ZZ-Mo edge, while the ZZ-S (ZZ-S-Mo) edge contained a few local grooves along with the terraces. Based on these observations, we propose a possible mechanism to explain the microscopic process of the oxidative etching from different edges on monolayer MoS$_2$, as shown in Fig. 3(b)–3(e). In general, the oxidative etching of MoS$_2$ can be expressed as the following reaction[33]:

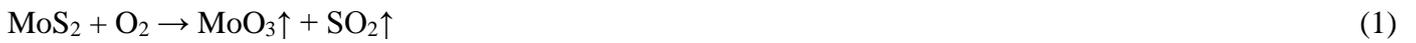

$$MoS_2 + O_2 \rightarrow MoO_3\uparrow + SO_2\uparrow \qquad (1)$$

As the ZZ-S-Mo edge and ZZ-S edge belong to the same crystalline orientation (-1010), and no obvious difference was observed in the residual morphology of these two edges after etching, we only focused on the case of the ZZ-S edge. As the oxygen molecules were preferentially chemically adsorbed onto the unsaturated S atoms on the edge, more S vacancies were formed and more unsaturated S atoms were exposed by the evaporated gas molecules. Such defects then became new starting points for the oxidative etching reaction. As the ZZ-Mo edge was energetically more stable, the reaction tended to continue along the S-terminated directions (marked by blue arrows in Fig. 3(b), 3(c)), leaving ZZ-Mo steps, which resulted in a groove morphology (blue dotted curves in Fig. 3(a)–3(c)). The observed grooves exhibited a relatively large curvature radius, rather than coned structures, which was probably a result of the local strain or the relatively higher reaction barrier at the Mo-terminated edges. Similar to previous reports [18-20], no MoO$_x$ particles were observed, indicating that all the reaction products evaporated in gas form. It can be inferred that the production gas was not the sublimated MoS$_2$, but MoO$_3$ and SO$_2$, since such etching cannot occur without oxygen[24]. However, in the case of the ZZ-Mo edges, oxygen molecules first bonded with unsaturated Mo atoms, forming Mo vacancies with exposed S atoms. The reaction then continued from generating unsaturated S atoms, owing to the preferentially crystalline orientation of the ZZ-S edge. Therefore, the etching direction was parallel to the ZZ-Mo edges (blue arrows in Fig. 3(e)). The atoms were peeled off layer by layer, forming relatively large atomic steps (yellow arrows in Fig. 3(d), 3(e)). Atomic steps with different heights were observed, since the absorption of oxygen molecules occurred randomly along the ZZ-Mo edge.

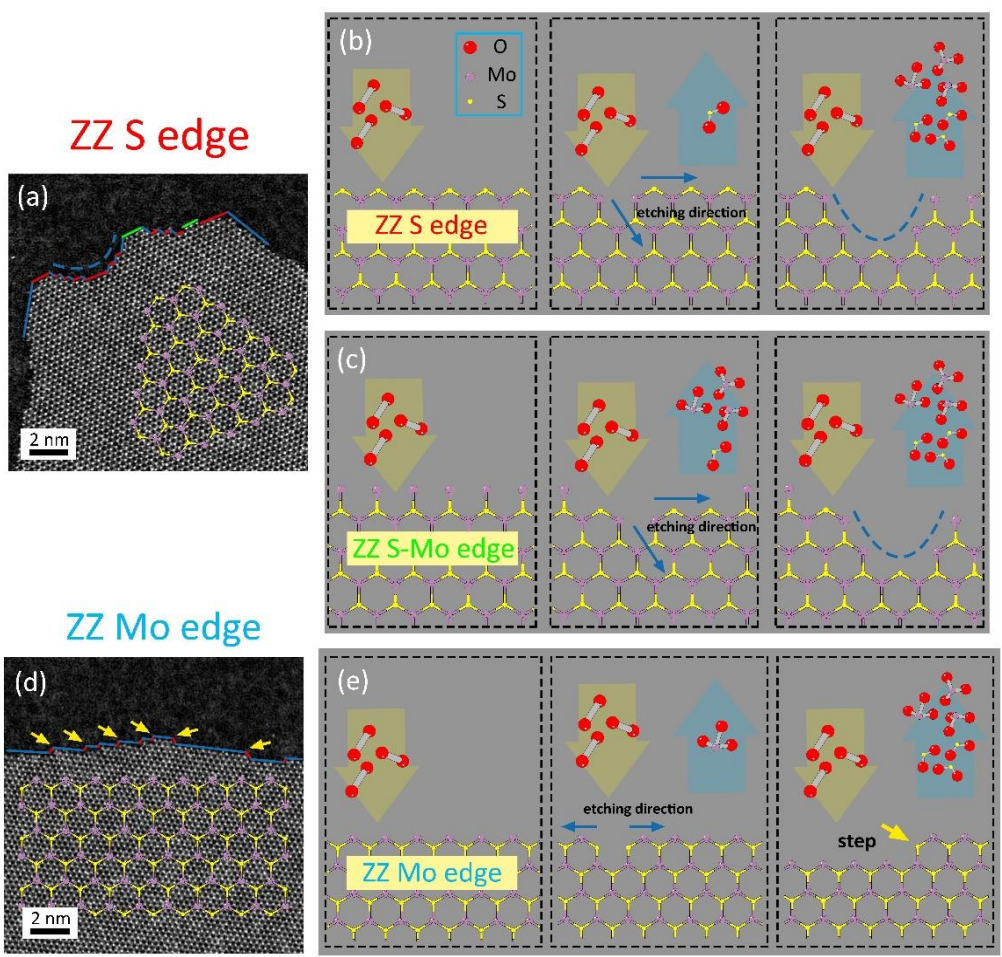

**Fig. 3** Etching mechanism schematic of different terminated edges in MoS$_2$. (a) ADF-STEM image of the etched zigzag S edge with three typical structures: zigzag Mo edge, zigzag S edge and zigzag S-Mo edge, denoted by blue, red and green lines, respectively. The hexagonal atomic model indicates the edge structures (purple: Mo, yellow: S). (b, c) Etching mechanism schematic of the zigzag S edge (zigzag S-Mo edge). Small blue arrows show the etching directions. (d) ADF-STEM image of etched zigzag Mo edge with two typical structures: zigzag Mo edge and zigzag S edge. (e) Etching mechanism schematic of the zigzag Mo edge. Small blue arrows show the etching directions and the yellow arrows indicate a formed atomic step.

Preferential etching was also observed along the grain boundaries of $MoS_2$ monolayers. In Fig. 4, the different domains are labelled with different colors, and they are also highlighted in the fast Fourier transform (FFT) using the same colors (original images shown in Fig. S5). The atomic resolution ADF-STEM images indicate that the oxidative reaction in the grain boundary (GB) started from the center to both ends, probably arising from the strain release at the ends of the GBs. Previous studies[23, 34] have indicated that GBs are generated from the reconstruction of dislocated cores when two domains merge with each other during growth, leading to plenty of unsaturated S atoms[35-36]. Therefore, atomic steps consisting of a ZZ-terminated Mo edge were still dominant at the residual edges, which indicated the same etching mechanism as that at a domain edge.

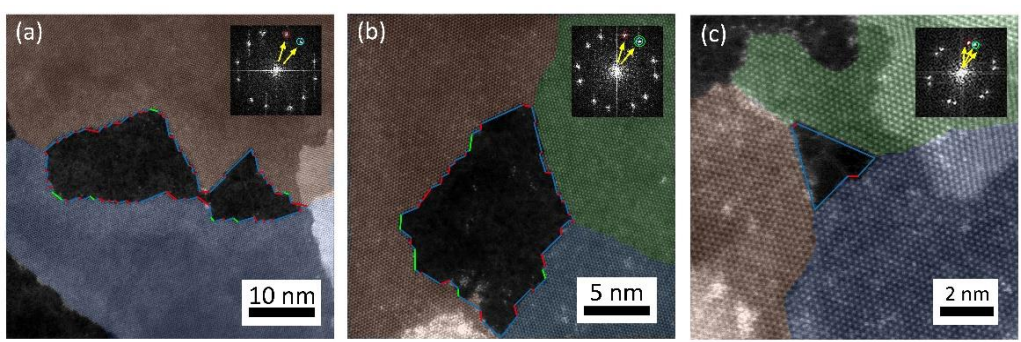

**Fig. 4** ADF-STEM images of etched grain boundaries in as-grown $MoS_2$ monolayers. The residual edges are marked with blue, red and green lines, representing zigzag Mo edges, zigzag S edges and zigzag S-Mo edges, respectively. Different grains are labelled by different colors, which are also marked with solid circles in the corresponding FFT patterns (insets).

Quantitative statistics were carried out to understand the relationship between the residual edge structure and etching position. From the histograms shown in Fig. 5, the ZZ-Mo edge (in blue) occupied above 70% of all the etching positions, suggesting it was relatively passivated in the oxidative etching reaction. Interestingly, when the etching occurred at the domain edges, there was almost no difference in the relative proportions of the ZZ-S edges and ZZ-S-Mo edges, while in GBs and inner domains, there were more ZZ-S edges than ZZ-S-Mo edges. The difference in structure where the oxidative etching originates could be a reasonable explanation for this observation. That is, after etching, both ZZ-S and S-Mo edges do not exist in long atomic step form, except for in the domain edge case, where the extra ZZ-S and S-Mo edges result from the original ZZ-S and S-Mo edges in the as-grown MoS$_2$ (in Fig. 1(b)). However, in the other two cases, most of the ZZ-S edges were counted as the corners of the atomic steps consisting of ZZ-Mo edges (yellow arrows in Fig. 3(d), 3(e)), in which quite a few ZZ-S-Mo edges were found.

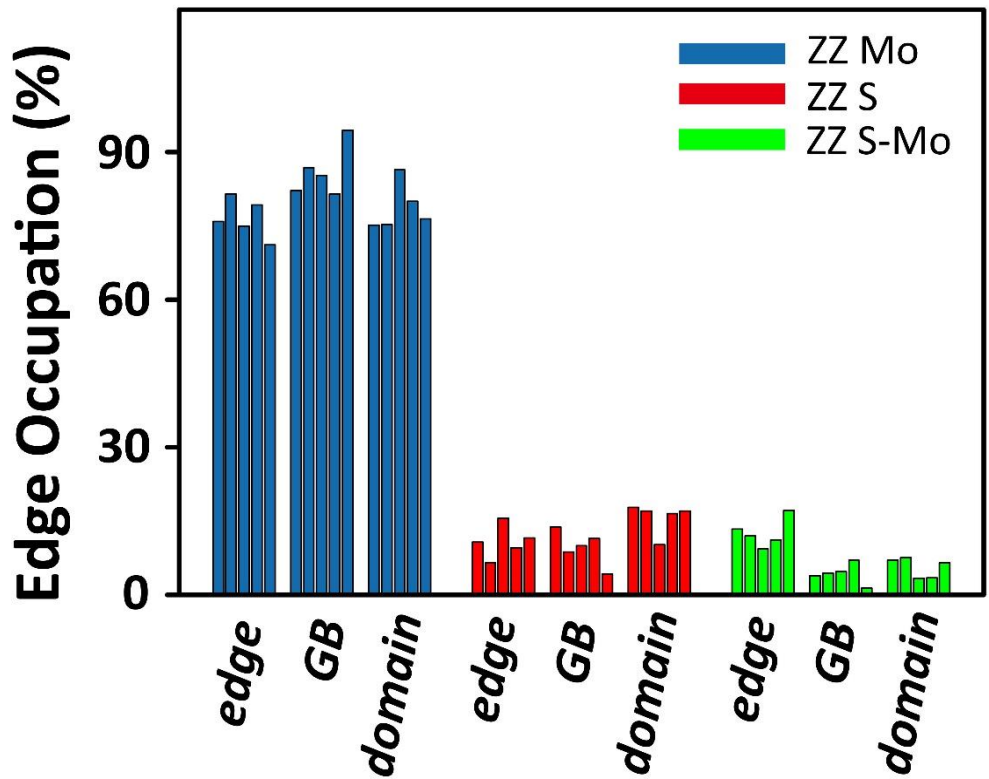

**Fig. 5** Statistics histogram of the residual edge structures after oxidative etching in different regions. Blue column: zigzag Mo edge; red column: zigzag S edge; green column: zigzag S-Mo edge.

An atomic process can be clearly inferred from our experimental results, that is, oxygen reacts first with unsaturated S atoms, which are abundant in GBs, ZZ-S edges and S vacancies, and unsaturated Mo atoms, rich in ZZ-Mo edges, then diffuses to the inner domain along certain crystalline orientations. Similar to graphite, two possible reaction routes are raised[37-39]: one is the oxygen gas directly reacts with the reactive defects sites on the basal plane of MoS$_2$ (Eley-Rideal (ER) mechanism), the other is the oxygen molecules are first chemisorbed and then migrate to react with the active sites on the MoS$_2$ surface (Langmuir-Hinshelwood mechanism). However, it still remains unclear if there are residual O atoms. As previously discussed, there were no clusters of MoO$_x$ observed from the oxidation, which also agreed with previous work. Atomic oxygen has been predicted to be thermodynamically favorable to replace at chalcogen vacancies or ZZ-Mo edges[17, 32], which was not obviously detected in our ADF-STEM characterization. And note that we did not find any clear evidence that the $MoO_x$ nanoparticles and PMMA residues around the $MoS_2$ flakes have any noticeable contrition to the oxidative etching. Electron energy loss spectroscopy (EELS) also failed to confirm the existence of atomic oxygen, owing to the limitation of both the electron microscope and samples: (i) The oxygen signal may be attributed to the residual $MoO_x$ during the growth or the residual PMMA during the transfer of graphene. (ii) Atomic resolution EELS is needed for the atomic oxygen at the edge or absorbed in S vacancies, which is far beyond the energy resolution of our microscope. (iii) High probe current and S/N ratio are also necessary, which may cause a relatively high damage rate for the 2D materials. The detailed study on the existence of oxygen requires further experimental investigations.

## 4. Conclusions

In summary, we have reported the "*in-situ*" oxidative etching on $MoS_2$ monolayers grown on a graphene substrate by precise positioning in TEM. The type and distribution of the residual edge structures in different positions have been analyzed based on the ADF-STEM images to further the understanding of the atomic mechanism of oxidative etching. Oxidative etching starts at defects and edges, and generates extra ZZ-Mo edges in the domains, which introduce a catalytic activity for HDS or HER into the inert basal plane of $MoS_2$. Knowing the mechanism of oxidative etching of TMD films will assist in improving the lifetime of devices and should facilitate their further application in electronics and catalysis.

**Acknowledgements**

We acknowledge financial support by the National Basic Research Program of China (No. 2014CB932500, No. 2015CB921004), National Science Foundation of China (No. 51472215, No. 51222202, No. 61571197 and No. 61172011) and the 111 project (No. B16042). This work made use of the resources of the Center of Electron Microscopy of Zhejiang University.

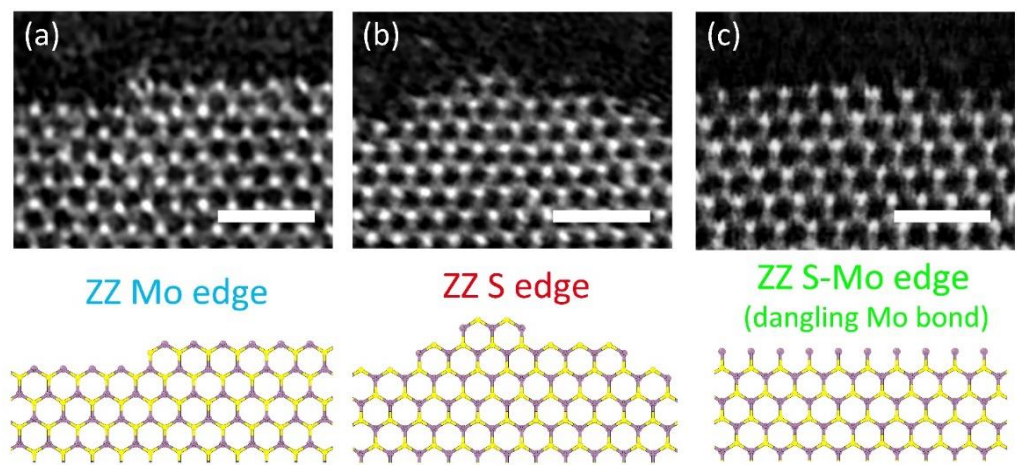

**Fig. S1** ADF-STEM images and structure models of three typical edge structures after etching. Scale bar: 1 nm.

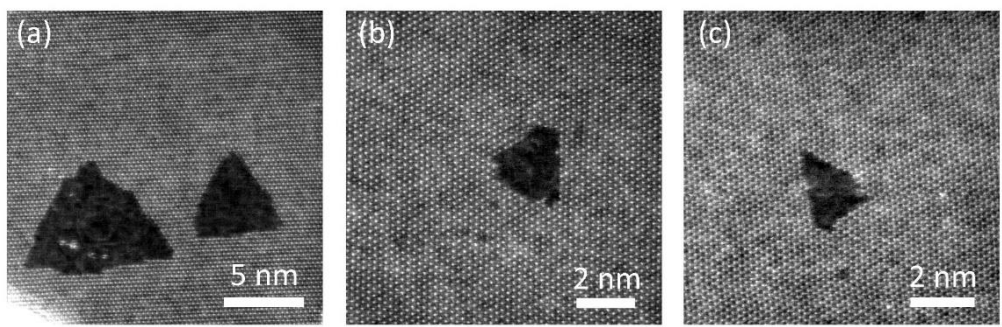

**Fig. S2** ADF-STEM images of triangular pits in the inner domain of monolayered $MoS_2$ on graphene substrate. The oxidation was carried out at 300 ℃ for 5mins.

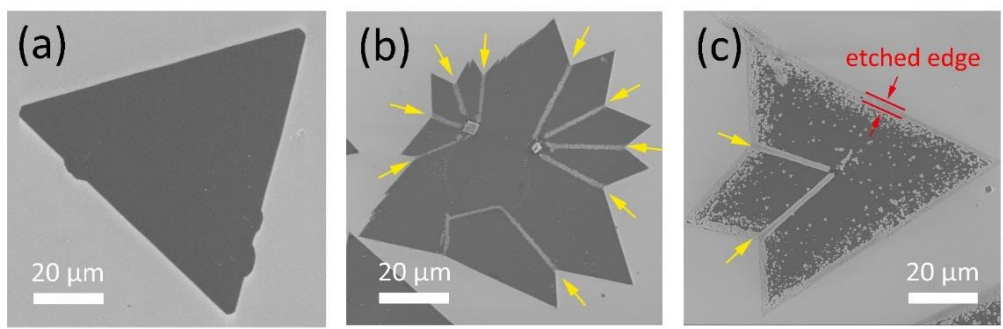

**Fig. S3** SEM images of oxidative etched $MoS_2$ monolayers grown on $SiO_2$/Si substrate: (a) 0 min; (b) 30 min; (c) 60min, all at 300 ℃. Yellow arrows shows the etched grain boundaries.

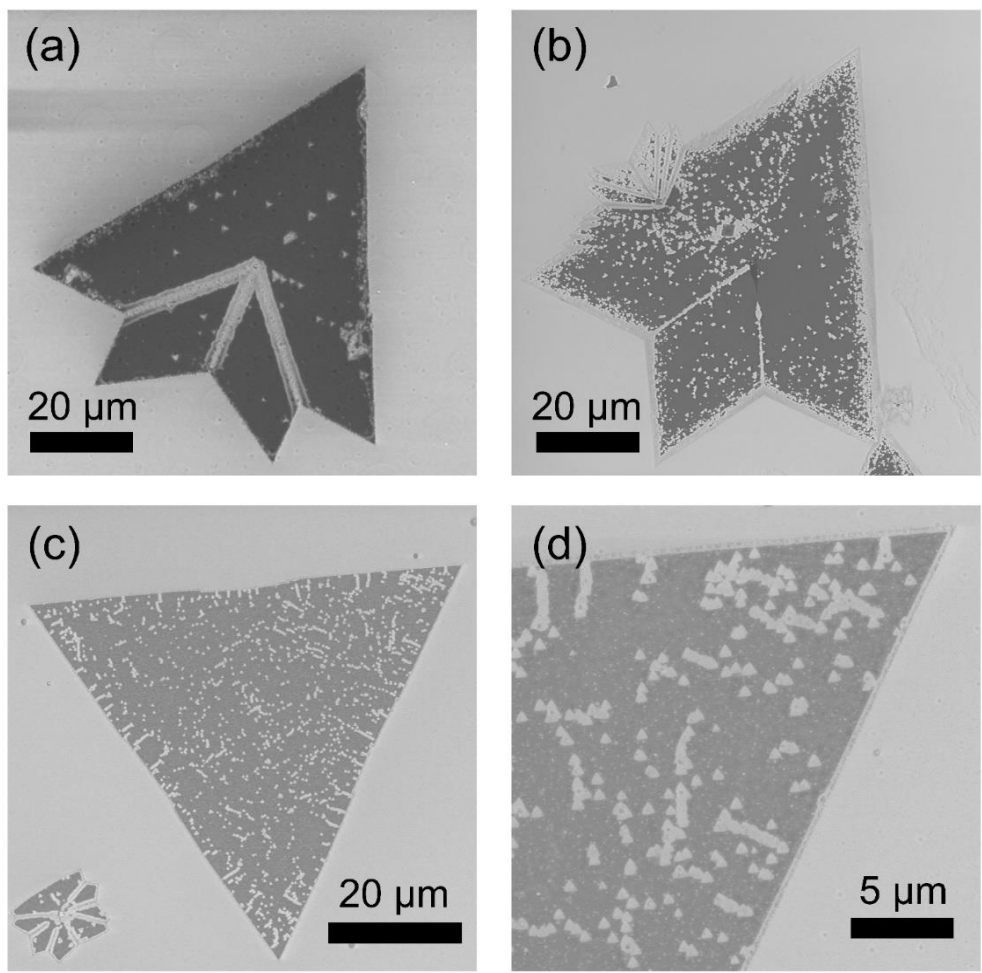

**Fig. S4** SEM images of monolayered MoSe$_2$ on SiO$_2$/Si substrate after oxidative etching. Oxidative etching was carried out at: (a) 325℃, 15 mins; (b) 325℃, 30 mins; (c) 325℃, 20 mins. (d) magnified image of (c).

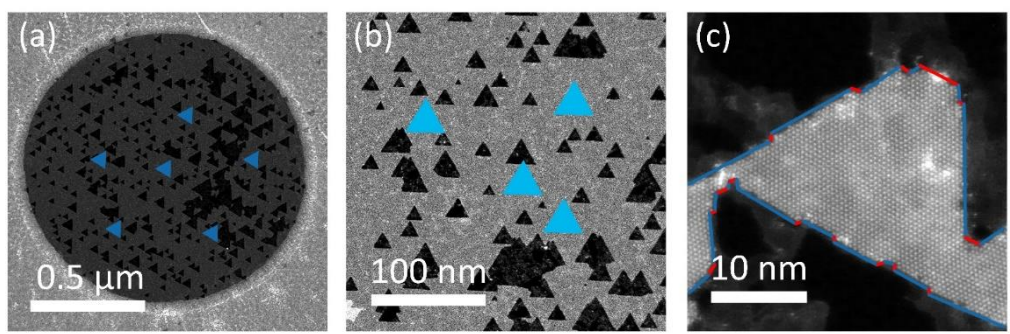

**Fig. S5** ADF-STEM images of oxidative etched MoS$_2$ monolayers transferred onto graphene/TEM grid. Blue triangles indicate the direction of the well-oriented etching pits.

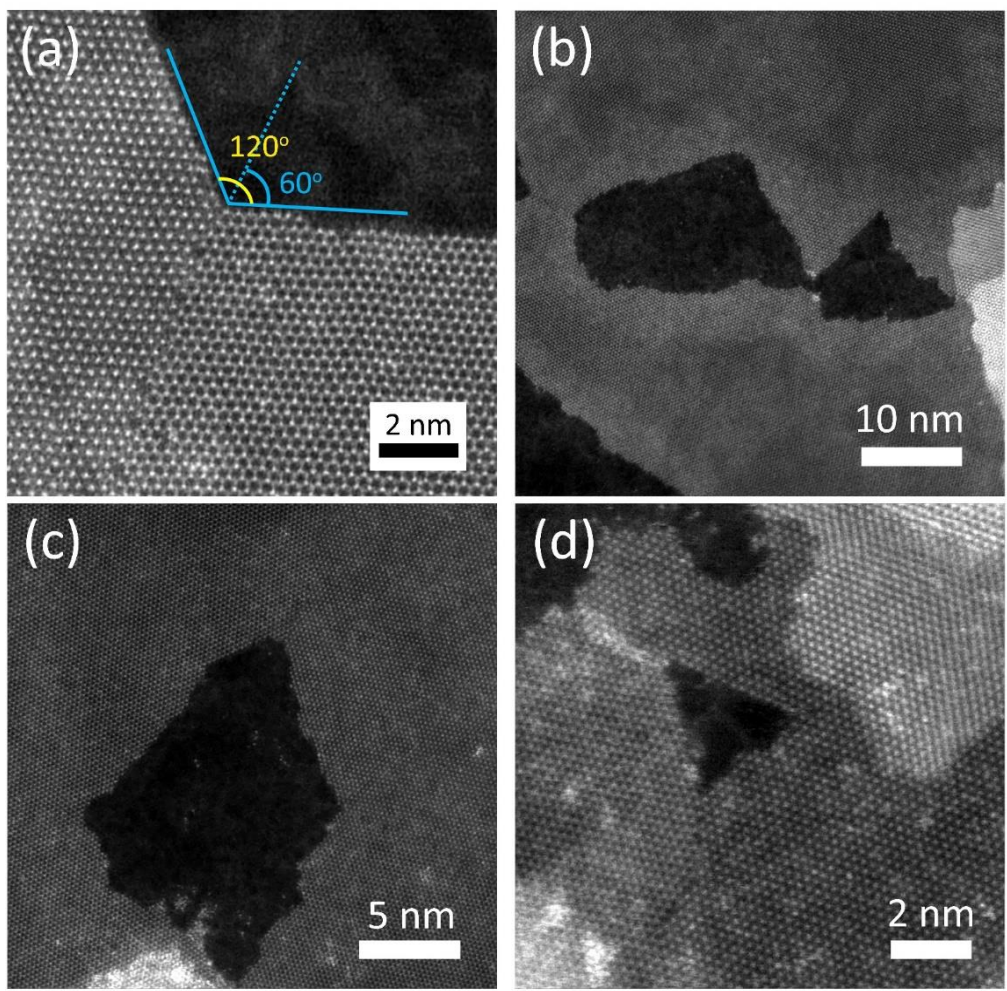

**Fig. S6** ADF-STEM images of grain boundaries of monolayered MoS$_2$ on graphene substrate. (a) ADF-STEM image of as-grown MoS$_2$ monolayer with 60º grain boundary. (b-d) ADF-STEM image of etched as-grown MoS$_2$ monolayer with grain boundaries.